# Silicon crystals for steering of high-intensity particle beams at ultra-high energy accelerators

A. Mazzolari[1], M. Romagnoni[1,2], E. Bagli[1], L. Bandiera[1], S. Baricordi[3], R. Camattari[1], D. Casotti[1,2], M. Tamisari[1,4], A. Sytov[1], V. Guidi[1,2], G. Cavoto[5,6], S. Carturan[8], D. De Salvador[8,9], A. Balbo[10], G. Cruciani[2], Thu Nhi Trans[11], R. Verbeni[11], N. Pastrone[12], L. Lanzoni[13], A. Rossall[14], J.A. van den Berg[14], R. Jenkins[15], P. Dumas[15]

[1] INFN Sezione di Ferrara, Via Saragat 1, 44122 Ferrara, Italy
[2] Università degli Studi di Ferrara, Dipartimento di Fisica e Scienze della Terra Via Saragat 1/C, 44122 Ferrara, Italy
[3] Dott. S. Baricordi, Via O. Putinati 67, 44123, Ferrara, Italy
[4] Dipartimento di Scienze Biomediche e Chirurgico specialistiche Università di Ferrara
[6] Università degli Studi di Roma La Sapienza, Piazzale Aldo Moro 2, 00185 Rome, Italy
[7] INFN Sezione di Roma, Piazzale Aldo Moro 2, 00185 Rome, Italy
[8] INFN Sezione di Legnaro, Viale dell'Università 2, 35020 Legnaro PD, Italy
[9] Dipartimento di Fisica e Astronomia, Università degli Studi di Padova, Via Marzolo 8, 35131 Padova, Italy
[10] Centro di Studi sulla Corrosione e Metallurgia "Aldo Daccò", Dipartimento di Ingegneria, Università di Ferrara, Via Saragat 4a, 44122 Ferrara, Italy
[11] European Synchrotron Radiation Facility (ESRF), 71 avenue des Martyrs, 38000 Grenoble, France
[12] INFN Sezione di Torino, Via Giuria 1, 10125 Torino, Italy
[13] Dipartimento di Ingegneria "Enzo Ferrari", Università degli Studi di Modena e Reggio Emilia Via Pietro Vivarelli 10, 41125 Modena, Italy
[14] Ion Beam Centre, School of Computing and Engineering, University of Huddersfield, Huddersfield, HD1 3DH, United Kingdom
[15] QED Technologies North America, 1040 University Avenue Rochester, New York 14607, United States of America

**Abstract** Experimental results and simulation models show that crystals might play a relevant role for the development of new generations of high-energy and high-intensity particle accelerators and might disclose innovative possibilities at existing ones. In this paper we describe the most advanced manufacturing techniques of crystals suitable for operations at ultra-high energy and ultra-high intensity particle accelerators, reporting as an example of potential applications the collimation of the particle beams circulating in the Large Hadron Collider at CERN, which will be upgraded through the addition of bent crystals in the frame of the High Luminosity Large Hadron Collider project.

## 1. Introduction

In a crystalline lattice atoms occupy well defined positions, resulting in regularly spaced distributions of electron and nuclei densities. This structure can be described as made by planes and rows of atoms. Exploiting this property, coherent scattering of electromagnetic radiation (e.g. X-rays) or of massive particle beams (e.g. neutrons, electrons, protons, etc…) by a crystal lattice occurs when properly mutually oriented.

Since the early '80s, bent crystals started to find application in accelerators as elements useful for beam steering [1-7] or splitting [8]. This was possible thanks to the availability of innovative ideas [7, 9, 10] and silicon crystals of sufficiently high crystalline perfection. Indeed, as a charged particle impinges on atomic planes or axis of a crystal at a sufficiently small angle, it is captured under channeling regime [11]. As the crystal is bent, its atomic planes become a pathway for propagation of the particle inside the crystal and then the deflection of particle beam occurs [5]. A first generation of experiments conducted in the '80, recorded deflection efficiencies in the order of few percent or less, which raised to few tens of percent in the '90s [5] mainly thanks to a proper design of the crystal geometry [12] and innovative crystal manufacturing approaches [13]. In more recent years, the discovery of numerous effects appearing whenever the crystal planes [14-18] or the crystal axes [19-24] are aligned to the charged particle beam direction was the results of reliable physical models, innovative crystal bending schemes [25, 26] and considerable improvements in crystal manufacturing techniques [27, 28]. The most recent results achieved at the Large Hadron Collider (LHC) [6] and the Super Proton Synchrotron at CERN [29-31] demonstrate that the technology readiness level reached by this technique makes it relevant for efficient deflection of particle beams in ultra-high-energy and intensity accelerators. Indeed, after a long R&D mainly carried in the frame of the UA9 collaboration, the use of bent crystals have been recently added as a part of the baseline upgrade of the collimation system of the LHC in the frame of the High Luminosity Large Hadron Collider (HL-LHC) project [32, 33] at CERN.

Bent crystals could be used also to extract the halo of the beam circulating in the LHC toward an extracted line with no cost for the collider-mode experiments, enabling a ground-breaking physics programme accessible within fixed-target experiments with the multi-TeV proton and ion beams [34, 35] of the LHC. While this proposal would result in a considerable effort, fixed target experiments using existing detectors and smaller changes to the LHC infrastructure have been suggested [36-38] for operations with the LHCb and ALICE collaborations. Operations with the LHCb detector are motivated by the aim to measure the electric and magnetic dipole moments of charmed baryons [37, 39-45], resulting in the determination of the magnetic moment of the charm quark. Within ALICE the use of bent crystals have been suggested in view of possible fixed-target experiments aiming to investigate physics related to the parton content of the nucleon at high-$x$, to the nucleon spin and to the quark-gluon plasma [46].

Crystal-based solutions [47] are also under consideration for the slow extraction of circulating beams in the Super Proton Synchrotron (SPS) at CERN [48] or at synchrotron facilities [49]. Aside from particle beam steering, crystals could also be used as innovative elements for generation of hard radiation [50, 51] or positron [52] beams and for the realization of compact photo-converter [8] and forward electromagnetic calorimeters [53, 54].

As compared to traditional magnetic optical elements used in accelerators, crystals operate without the need of a power source, generally do not require a cryogenic environment, and are extremely compact and light-weight, typically less than 1 Kg, making them appealing for integration in particle accelerators. The use of crystals have been suggested also for the Future Circular Collider (FCC) [39, 41, 55-57], the International Linear Collider (ILC) [52, 58], the Compact Linear Collider (CLIC) [58, 59], and the Circular Electron Positron Collider (CEPC) [60] and it could be exploited in a future muon collider design [61, 62]. Given the extremely high intensity and energy reached in modern and future particle accelerators such as the HL-LHC or in future planned ones, the implementation of bent crystals at these facilities demands facing various technological challenges and a considerable improvement of the state-of-the-art in the manufacturing of crystals for scientific or technological application.

As an example of potential application, in this paper we focus our attention mainly on the solutions for manufacturing and characterization of crystals with a geometry optimized for the collimation of the LHC ion beam in the framework of the HL-LHC project. The successful development of such crystals is based on a merging of ultra-modern technologies used in microelectronics, X-ray science, ultra-precise optical and mechanical machining and might open innovative schemes for particle beam deflection at future accelerators. For example, the geometry of crystals suitable for collimation [63, 64] or extraction [37, 38] of the LHC circulating beams are similar, so a revisitation of the crystals developed for the collimation of the LHC beam might enlarge the possibilities for fixed-target experiments at the LHC.

## 2. Bent crystals as particle beam collimators

A reliable collimation system is a key component of any particle accelerator, especially for ultra-high energy/intensity colliders based on superconducting magnet. In particle accelerators, indeed, various mechanisms can cause particles of the beam to enter into unstable orbits, causing the formation of a beam halo and the growth of the beam emittance. This turns into uncontrolled beam losses, potentially causing damage to the accelerator or other perturbations to the operation such as activation of the accelerator components, increased background to the experiments, the quenching of superconducting magnets, diffuse radiation damage, etc… The main purpose of a collimation system is to safely dispose of beam halo losses: a classical solution to this problem consists in the adoption of a multi-stage collimation system, where a sequence of targets intercepts the halo of the beam. The system effectiveness is determined by the interaction between the beam particles and the collimator active material, i.e. the electromagnetic, elastic and inelastic nuclear interactions. While elastic interactions do not change the structure of the target or of the particle, inelastic interactions may result in multi-particle final states (with the relevant case of single or double diffractive events). These reactions may result in deleterious effects to the accelerator components, such as targets, detectors, collimators, and the general accelerator environment and could represent a relevant issue especially for accelerators such as the HL-LHC or the FCC, where extremely high beam intensities and energies are foreseen. The concern related to quenches of cold magnets is particularly important for the LHC and determines the most challenging requirement for the LHC collimation system, which demands a cleaning efficiency larger than 99.99%.

In the present LHC collimation scheme (see figure 1), a "primary collimator" intercepts the beam halo. A fraction of particles is absorbed in the collimator, while the remaining produces a "secondary halo" and a spray of unwanted "new" particles emerging as a result of single or double diffractive collision events. The particles that escaped from the primary collimator are partially intercepted by secondary collimators, exhibiting a similar behavior to the primary one. If needed, additional collimators are installed after the secondary one. The overall result of this collimation chain is to absorb the beam halo and the method constitutes a classical solution at various accelerators [65-69]. This approach might not provide sufficient cleaning in operations with future accelerators, such as the HL-LHC at CERN, where the stored energy will approach 700 MJ per beam [32, 70]. In particular, for the case of operations with heavy ions, the collimation efficiency degrades as a consequence of the large cross-section of fragmentation processes, resulting in isotopes of different rigidities. Often, these ions do not receive a sufficiently large angular kick to be intercepted by the secondary collimators and they deviates from the main beam. Such lost ions could cause a superconducting magnet to quench, or in the worst case, damage the hardware of the accelerator.

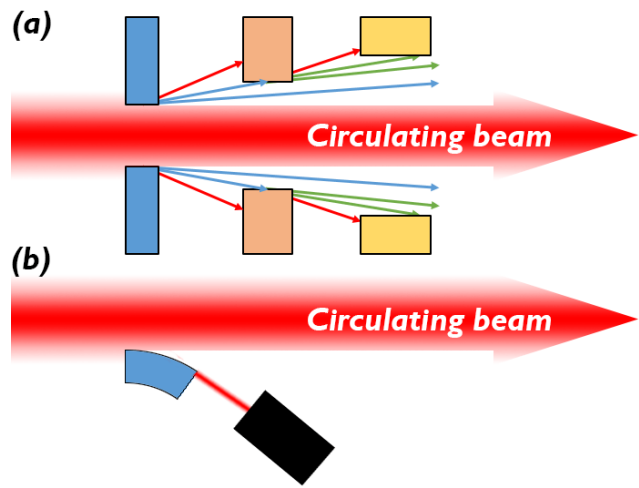

Figure 1 (a) The classical approach for the collimation of circulating beams in a particle accelerator is based on a chain of materials placed at different apertures along the beam path. Interaction between the "primary halo" and a first block of material (blue box) leads to partial absorption of the halo and, as a by-product of the interaction, to a spray of particles (secondary halo) made primary of halo particles (red arrows) which are not stopped and "new particles" generated in single or double diffractive events (green or blue arrows). The secondary halo is partially absorbed by a second block, which plays the same role for the secondary halo as the first block for the primary halo. (b) Collimation based on a bent crystal as primary collimator. A bent crystal (blue) is aligned to the primary halo and channels its particles between the crystal atomic planes. Channeled particles are efficiently deflected to a massive absorber (black). Given the nature of interaction between the crystal and the beam, the rate of inelastic interactions is strongly suppressed. The elements depicted in this figure have sizes that are not to scale. The figure is inspired to figure 5-18 of chapter 2 of [71].

The regular distributions of atoms in a crystals lattice make them suitable targets for collimation purposes. This results have been already achieved at RHIC in the collimation of circulating ions beams [72] and have been suggested for the collimation of the LHC in [64] and subsequently studied in more detail in [63]. When the atomic planes of a crystal are not aligned to the high-energy particle beam, interaction between the beam and the crystal is very similar to the interaction that the beam would experience in an amorphous target [73]. As the crystal is aligned to the particle beam in order to excite channeling [5], beam-crystal interactions show unique features. Indeed, under this circumstance each particle of the beam interacts with the electric field naturally present between atomic planes. This interaction drastically influences the motion of each particle of the beam, which passes from a "random" motion for the case of non-alignment, to an oscillatory motion between two neighboring atomic planes for the case of alignment [5]. For example, in case of interaction with a silicon crystal, the wavelength of the oscillation is ~56 µm if the channeled particle energy is 400 GeV, and rises up to ~235 µm at 7 TeV. Therefore, particles move across the crystal in regions of low-atomic density. Crystals whose geometry is optimized to maximize beam steering lead an efficiency up to ~83% for operations in "single-pass" mode [15, 18], and up to values close to 100% thanks to the multi-pass mechanism [10] in the case of operations in circular machines [12].

Assuming that a sufficiently robust absorber to dispose the extract beam is available, with respect to the classical collimation schemes, a scheme based on crystals might deliver two potential advantages:

1) As a crystal is aligned to channel particle beams between its bent atomic planes, the crystal acts as a low-loss waveguide for the beam [74, 75], resulting in a strong suppression of fragmentation [76, 77] and nuclear interactions with respect to the case of non-alignment and with respect to the case of traditional schemes.
2) The reduction of the machine electromagnetic impedance. While the classical collimators are meter-long bulky objects and many replica of them are needed, only one single crystal collimator as short as a few mm can be used, followed by a single – although more massive than standard collimators – absorber (per beam and per collimation plane: horizontal and vertical).

### 3. Crystal manufacturing and characterization

The manufacturing of bent crystals and benders for applications in particle accelerators requires a highly multi-disciplinary approach, with the merging of techniques typically used for silicon micromachining, ultra-high precision optical and mechanical machining, the most advanced X-ray characterization techniques, and capabilities in ultra-precise metrology.

In the following, we discuss the features of the crystals which might influence steering efficiency in modern particle accelerators, focusing our attention in particular on the manufacturing of crystals that could be suitable for collimation of the HL-LHC proton or ion circulating beams.

### 3.1. Starting material: silicon wafers of ultra-high crystalline quality

Crystals are usually manufactured starting from 100 mm diameter silicon wafers a few mm thick (typically 2 mm) and (110) oriented. The orientation of the largest faces of the wafer determines the orientation of the crystalline planes that will interact with the particle beam. Channeling efficiency and the rate of inelastic interactions [75] are mainly influenced by the distance between atomic planes and their atomic density, as a result the most performing planes are the (110) and the (111) [5]. The average distance between (111) planes (1.568 Å) is smaller than for (110) planes (1.92 Å), consequently the steering efficiency of the latter is higher and the rate of inelastic interaction with the lattice is lower. To confirm this statement, in [75] the performance of crystals with thickness and bending angle suitable for the collimation of the HL-LHC circulating beam were compared in terms of the ratio between the rate of inelastic interactions occurring as the crystals were oriented and not oriented to excite channeling of a 400 GeV proton beam. As (110) crystal were oriented for channeling, nuclear interactions reduced to only ~27% with respect to a condition of non-alignment. For crystals offering (111) channeling planes the reduction leveled to ~36%.

To aid the further manufacturing steps, wafers with "miscut angle" (the angle between the physical surface and atomic planes of the crystal) less than or equal to 200 µrad were selected from a large stock of wafers. From this subset, wafers of highest crystalline perfection were selected.

Among the various crystallographic defects that might be present in a crystal, dislocations play the most detrimental role for the channeling efficiency. Presence of this defect results in a deformation field that propagates in the crystal up to large distances from the location where the dislocation appears. A particle that is initially channeled, interacting with the deformation field induced by a dislocation, would suffer "immediate" dechanneling. In order not to degrade the channeling efficiency, a dislocation density lower than $1/cm^2$ is required [78]. With the aim to satisfy the requirement on what is nowadays considered as an extremely low density of dislocations even for microelectronics (where a level of 50 pits/$cm^2$ is accepted), a large stock of wafers was processed with chemical etchings capable of highlighting the presence of single dislocations [79, 80]. The wafers that satisfied the requirement of less than one dislocations/$dm^2$ were further checked by means of X-ray topography techniques [81] at the European Synchrotron Radiation Facility (ESRF). Both the techniques have a sensitivity of 1 dislocation over many $cm^2$. The wafers with dislocation density lower than $1/dm^2$ were further processed.

### 3.2. Crystal orientations and miscut angle

Interaction of crystals with circulating beams exploits a few-micron-deep crystal portion, close to one of its surfaces [63]. Depending on the dynamics of a circulating beam, the angle between the optical surface of the crystal and its atomic planes ($\Theta_m$ in figure 2b) might play a role in determining the channeling efficiency of the beam between the atomic planes of the crystal [29, 30, 82, 83]. That angle is typically identified as the "miscut" or "off-axis" angle: it is mainly generated in the process of slicing the silicon ingots into silicon wafers, as a result of mechanical tolerances of the manufacturing equipment .

As shown in figure 2, any residual miscut imply that particles approaching the surface of the bent crystal with a small impact parameter may be only partially deflected, unable to reach the secondary absorber and possibly lost in sensitive area of the accelerator.

The miscut angle must be kept as small as possible, and in any case should be much smaller than the angle subtended by the bent atomic planes. For example, collimation of the LHC circulating beam requires a crystal with a bending angle of 50 µrad [63], leading to an ideal miscut angle a few orders of magnitude lower than what is routinely achieved for microelectronics and X-ray synchrotron facilities, both define the quality standard for the technological and scientific communities operating with silicon crystals.

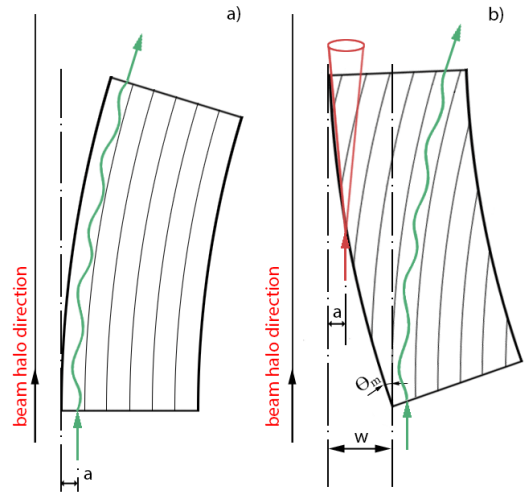

Figure 2: (a) Sketch of a bent crystal with zero-miscut angle aligned to the halo of a beam circulating in a particle accelerator. Particles (green trajectory with an arrow) are captured in the channeling regime independently of their impact parameters to the crystal, $a$. (b) A miscut angle, $\Theta_m$, introduces, a correlation between the impact parameter and the crystal-beam orientation. Only particles with sufficiently large impact parameter, $w$, are captured under channeling regime and then steered. Particles with lower impact parameter (red arrow) interact with the crystal as if it were an amorphous target and their direction is spread out by multiple Coulomb scattering (red cone).

Challenges in manufacturing of mm-size crystals with an ultra-small miscut comes from the need of an ultra-accurate characterization setup and, mostly, from the availably of a polishing approach which accounts for unavoidable deformation of the crystal as a result of its fixing to the polishing equipment. To adjust the miscut, ultra-precise polishing techniques typically used for the manufacturing of ultra-precision optics were revisited to operate on silicon. The wafer's largest surfaces were indeed polished by means of a Magnetorheological Finishing (MRF). MRF is a precision polishing method, mainly developed by QED Technologies (QED) [84, 85], to overcome many of the fundamental limitations of the traditional optics finishing. Conventional optical polishing uses stiff full-aperture laps to smooth (to improve the roughness) and to reduce the overall form error. While this process can work extremely well in delivering extremely flat or perfectly spherical surfaces, it fails as the surface shapes become more complex, e.g. aspheric or free-form, and hardly allows a precise control in the tilting of the sample to be polished. On the other hand, MRF is a deterministic finishing process, i.e. a process where the amount of material removal occurs in a highly predictable fashion, thus leading to high convergence rates and accurate estimates of the cycle times [86]. The MRF process is based on the use of a magnetorheological fluid, whose unique property is that its viscosity changes by several orders of magnitude when it is introduced into a magnetic field, essentially turning from a liquid to a quasi-solid in milliseconds.

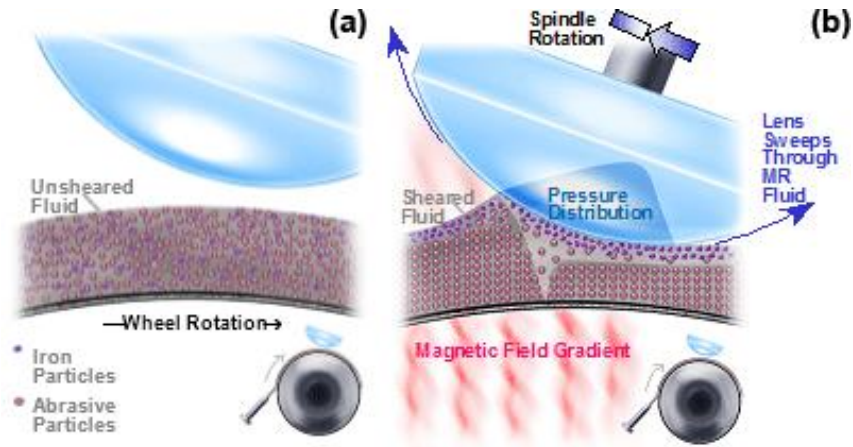

Figure 3 (a) Creation of MRF polishing "spot". When the magnetic field is off, there is a random distribution of iron abrasive particles in the ribbon of fluid being transported by the rotating wheel. (b) When the magnetic field is turned on, the iron particles align and form chains giving to the fluid structure and stiffness. In addition, the water and the abrasives material move to the surface because the iron particles are attracted toward the wheel. When the workpiece is inserted into the fluid, the converging gap creates a highly sheared fluid layer that removes material with very low normal forces acting on the individual abrasive particles.

The forces acting on the surface are predominantly tangential [87, 88]. The normal forces on the individual abrasive particles are very small (limited to hydrostatic and kinetic). This contrasts with conventional polishing techniques where an abrasive material is forced into the surface through the action of a lap (either bound or loose). In that case normal forces can dominate, creating scratches, sub-surface damage, and stress.

As a first step, the wafer surfaces were polished to reduce their flatness from a typical value of a few μm to less than 0.05 μm over their full area: surface topology was characterized with a Zygo 6" Verifire AT 1000 interferometer operating in Fizeau configuration. Subsequently, $\Theta_m$ was characterized by means of a high-resolution X-ray diffractometer (Panalytical X'Pert³ MRD XL) following the approach described in [89], and its value was decreased with successive polishing steps of MRF with miscut measurements. A record $\Theta_m$ value of less than 2 μrad was achieved after a total of 2 iterations.

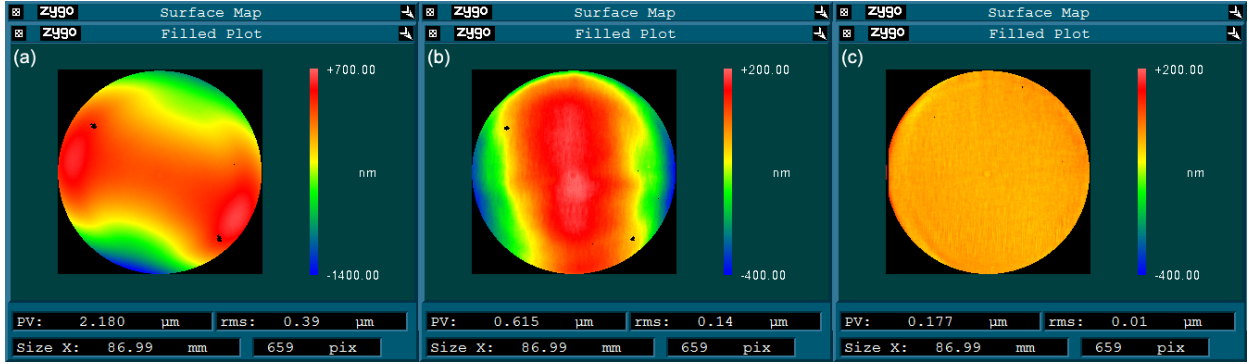

Figure 4 (a) Interferometric characterization of a wafer surface prior to the polishing operations shows surface flatness of 0.39 µm. (b) After a first polishing run with a high removal rate MRF fluid, the surface flatness was improved to a value of 0.14 µm. (c) After a second polishing run using a low roughness fluid, surface flatness improved to 0.01 µm. Black dots in (a) and (b) are fiducials used to orient the wafer in the measuring step.

### 3.3. Schemes for crystal bending

The bending angle that the crystal must impose on the particle beam is determined by the needs and constrains of the experimental setup. In practical applications, it may vary from a few to some tens of µrad, for example for the case of collimation [63] or extraction of the LHC [34, 37] or FCC [55] circulating beam, to more than 10 mrad for the case of experiments aiming to perform fixed-target experiments at the LHC [39-44, 90].

Various approaches have been developed to achieve the needed deformation state: the most investigated are based on the action of a mechanical bender imparting the wanted deformation to the crystal. Alternative approaches are under investigation: they are based on the deposition of thin (few tens of nanometer thick) films generating stress in the crystal [91] or machining of a crystal surface to generate a superficially thin damaged layer under controlled conditions (ion implantation [92], sandblasting [93, 94], grinding [95], surface grooving [96, 97]). Independently from the method used to deform the crystal, it is important to avoid unwanted deformations of the crystal itself. In particular, bending angle of the crystal and torsional deformation (see paragraph 3.1) must satisfy tolerances which are typically very tight. Moreover, in case of implementation of the crystal in a setup where the desired level of vacuum is achieved through thermal bake-out cycles (such as in the LHC), the deformational state of the crystal must be stable against such thermal cycles.

In the following, we will focus on bending approaches based on the use of mechanical benders, as this approach is nowadays the most experimentally investigated.

For the case of experiments related to fixed target experiments in the multi TeV regime, crystals capable of bending angle of various mrad are required [39-44, 90]. This, jointly with the fact that at that scale energy channeling becomes inefficient at bending radii below few meters [5], results in crystals of a few cm thickness. Under such circumstances, crystals can be bent exploiting schemes which impart a "primary curvature" to the crystal. Typically, the crystal is clamped between the surfaces of a properly machined jaw (see paragraph 4.1).

The case of applications expecting the interaction of a crystal with the halo of a circulating beam, such as beam collimation or extraction, requires crystals properly designed to reduce as much as possible beam losses. Compliance with this requirement typically results in crystals of thickness of some mm with bending radii of curvature larger than a few tens of meters. For such a case, exploitation of schemes based on a "primary curvature" are not feasible. Crystal bending relies on the use of secondary or tertiary deformations, arising because of a primary deformation imparted to the crystal through a mechanical bender. Two schemes are typically used: we consider a bar of an anisotropic material such as silicon, deformed under the action of moments (*M*) acting at both its ends (see Figure 5). Displacement field along the *x*, *y* and *z* axes turns out to be [98]:

$$\begin{cases} u = \frac{M}{2I}(2a_{13}xy + a_{36}y^2 + a_{35}yz) \\ v = \frac{M}{2I}(-a_{13}x^2 + a_{23}y^2 + a_{33}(hz - z^2) - a_{35}xz) \\ w = \frac{M}{2I}(a_{35}xy + a_{34}y^2 + a_{33}y(2h - l)) \end{cases} \quad (1)$$

where $a_{ij}$ are the components of the compliance matrix of the crystal and *I* is the moment of inertia of the cross sections with respect the *x* axis.

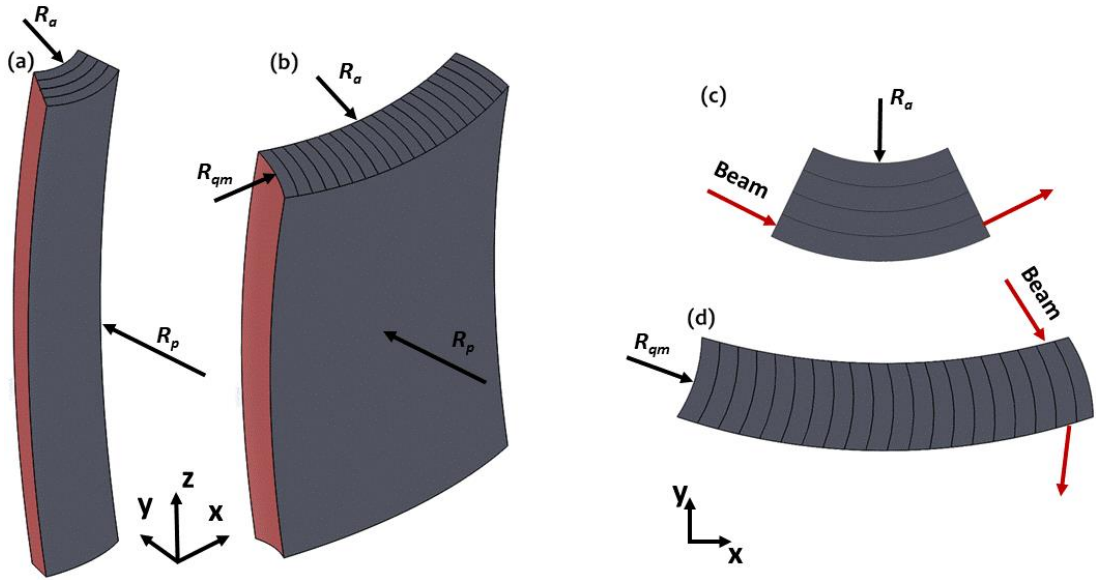

Figure 5: (a-b) silicon crystals bent under bending moments (not shown) acting at their ends. (b-c) cross sections of the crystals taken at their middle along the *z* axis. The crystal sketched in (a-c) exploits anticlastic deformation manifesting along the *x* axis, occurring for any crystallographic orientation of the crystal. The crystal sketched in (b-d) exploits instead the so called "quasi-mosaic" deformation, occurring only under a proper choice of crystallographic orientations as a bending of the crystal planes along the smaller dimension of the crystal (in this case along the *y* axis).

The dependence of $v$ in Eq. 1 over $x^2$ shows that after the deformation, the crystal assumes a saddle-like shape, characterized by a "primary curvature" imparted by the mechanical bender, and a secondary curvature, owing to the "anticlastic deformation", characterized by an "anticlastic radius". The ratio between the "anticlastic radius" and the "principal radius" corresponds to the Poisson ratio of the crystal (i.e. $a_{33}/a_{13}$) in the XZ plane. Being the Poisson ratio of any material between -0.5 and 1, adjustment of principal radius of the crystal provides a fine adjustment of the anticlastic deformation. Moreover, a proper choice of crystallographic orientations leading to $a_{36} \neq 0$, thus resulting in a tertiary deformation, typically identified as "quasi-mosaic" [99] deformation (see Figure 5), occurring as a bending of the planes along the crystal thickness.

The choice of bending mechanism influence the possible choices of the channeling planes. While anticlastic deformation manifests for crystals of any crystallographic orientation, quasi-mosaic deformation permits exploiting only (111) planes (or planes of higher Miller indices which would deliver low steering efficiency [100]). Moreover, quasi-mosaic deformation does not allow to exploit axial channeling, which might be useful for future studies aimed to extraction or collimation in circular accelerators [101]. Due to such limitations, crystals exploiting anticlastic deformation are preferable.

3.4. Crystals shaping

Shaping of a crystal typically relies on a proper matching between chemical and mechanical methods, which we borrow from approaches well established in the field of silicon micromachining, and which we adapted to the manufacturing of macroscopic silicon crystals.

The wafer previously processed was diced with a high precision dicing saw (Disco DAD 3220) to crystals whose lateral sizes are about ~0.1 mm wider than the final wanted size. Dicing parameters in terms of rotational speed, feed rate, blade grit, and thickness have to be optimized to reduce subsurface lattice damage induced in the crystal by the dicing operations as much as possible. Nevertheless a damaged layer extending for a few microns below the diced surface [27] is left after dicing. To recover the crystalline quality, purely chemical, chemical-mechanical polishing, or a mix between those techniques can be adopted to recover the crystalline quality of the surfaces which will be the entry/exit faces for the beam (chemical-mechanical polishing is employed as a last step for such situations were mirror-like surfaces are needed). At the same time, crystal sizes are reduced from the 0.1 mm in excess.

For the case of crystals for collimation of the LHC beam, an optimal thickness of 4 mm and a bending angle of 50 µrad were selected [63]. After being polished to decrease its miscut, a 2 mm thick wafer was diced to strips of 4.1x55 mm$^2$ lateral sizes. The lateral faces (sizes 2x55 mm$^2$) of the strips are parallel to the (110) face to within 1 degree. Subsequently to dicing, those faces were chemo-mechanical polished to remove the sub-surface lattice damage and at the same time reduce crystal size along the beam from ~4.1 to 4.00±0.01 mm.

3.5. Characterization of crystalline quality

A crucial aspect that crystals for applications in high-energy accelerators must satisfy is the absence of an amorphous or a crystalline-damaged layer on the crystal surface parallel to the direction of beam propagation. Indeed, the impact parameter of the beam with respect to the crystal is as small as a few μm [63]. Therefore, the crystalline quality of the surfaces hit by the beam is important to ensure a maximum channeling efficiency. Channeling must indeed occur starting from the very first atomic layers of a crystal [102, 103].

With the aim to characterize crystalline quality of the produced surfaces, an extensive experimental program was carried out using the most advanced methods typically used in semiconductors manufacturing.

Various authors already characterized subsurface damage and stress in amorphous materials treated by MRF [87, 104, 105]. However, a reliable and deep study of the crystalline quality of MRF-treated crystals through analysis techniques typically used to characterize crystalline materials is still missing. In particular, it should be pointed out that for a crystalline material there is no correlation between surface roughness and crystalline quality of the treated surface.

At first, high-resolution X-ray Bragg diffraction was used for a preliminary characterization of crystal surfaces in terms of dislocations or residual strain. As illustrated in figure 6, Bragg diffraction occurs when radiation, with a wavelength comparable to atomic spacing, is scattered in a specular fashion by the atoms of a crystal and undergoes constructive interference.

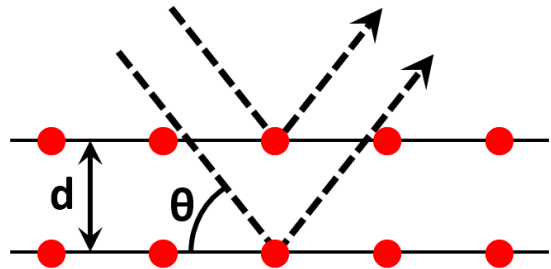

Figure 6: two X-ray beams with identical wavelength and phase approach a crystal and are scattered off two different atoms. The lower beam traverses an extra length of *2dsinθ*. Constructive interference occurs when this length is equal to an integer multiple of the wavelength of the radiation. Such condition can be satisfied properly rotating the crystal with respect to the incident X-ray beam.

High-resolution diffraction of X-rays is accomplished using a highly collimated and monochromatic beam and is routinely used worldwide to investigate deviations of a crystalline structure from an ideal crystal, which can be induced for example by crystallographic defects, strain or mosaicity. This technique foreseen orienting the crystal to the beam in such a way as to excite X-ray diffraction and counting the number of diffracted x-rays while rotating the crystal around such position. Indeed, accounting for propagation and absorption of x-rays by the crystal, it is expected that the diffraction occurs within a certain angular range. Recorded diffraction profiles (see figure 7) are typically called "rocking curves": a proper study of the profiles of such curves in terms of the dynamical theory of x-ray diffraction [106] reveals important information related to the perfection of crystal. For example, broadening of the rocking curve typically arises from a strained lattice, while a too high counting rate at the tails is a typical indication for the presence of dislocations.

A state of the art X-ray diffractometer (Panalytical X'Pert³ MRD XL) operating with a Cu anode source installed at INFN-Ferrara was used to perform high resolution X-ray diffraction characterization. In this machine, the X rays generated from an X-ray tube are collected by a Göbel mirror [107], which delivers a beam with a divergence of about 0.02°. X-rays coming from the mirror are injected in a compact monochromator exploiting four reflections on (220) oriented Ge crystals, then conditioned to a size of 5x0.2 mm². The beam output from the monochromator has a typical divergence of 0.003° and a wavelength relative spread less than $2*10^{-5}$ and is used to probe the crystal surfaces in terms of the presence of dislocations, lattice strains and relatively thick amorphous layers. Figure 7 shows the X-Ray rocking curves of (220) plane, i.e. the second order diffraction from the (110) lattice plane, obtained on both the surfaces of the silicon crystal parallel to the particle beam and on the surface which represent the entry face for the beam, compared to the rocking curve recorded on a reference crystal. A comparison of these rocking curves demonstrates an extremely high quality for both the surfaces subjected to MRF and chemo-mechanical polishing. The presence of a number of dislocations in excess over the reference sample, possibly induced by the surface machining, would be revealed by the scattering tails around the main peak induced by the dislocation's deformation field. As can be noticed no differences with respect to the tails of a reference crystal is detectable. Moreover, also the full-width-half-maximum of the rocking curve (~0.003°) coincided with the values expected for a case free from defects such as strains or mosaicity.

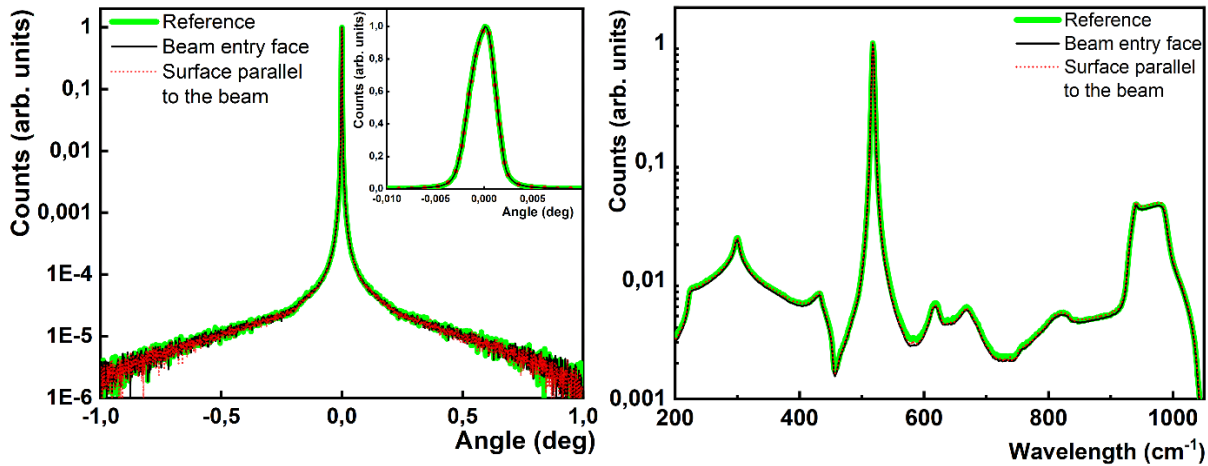

Figure 7. (a) High resolution X-ray diffraction rocking curves recorded on the surface of the crystal parallel to the propagation of the beam (red dashed curve), on the beam entry face (black curve), and on a reference surface (green curve). Absence of deviation of the tails from an ideal profile is indication of absence of dislocations. The inset shows a zoom of the rocking curves in the central region. Full width half maximum of the rocking curves are identical to the ones of a reference, indicating the absence of strained layers (b) Micro-Raman spectra collected on the crystal surface parallel to the beam (red dotted curve) and on the beam entry face (black line) are compared to micro-Raman spectra recorded on a reference sample (green). The crystal surface results to be free from strain or amorphous components, which would appear as additional peaks.

Operating with an 8 KeV X-Ray beam, we obtained a signal which is averaged over ~13 µm below the crystal surface. To obtain information from a thinner region below the surface, we used micro-Raman spectroscopy [108, 109]. This technique relies upon inelastic scattering of photons and is typically used to determine vibrational modes of molecules or atoms in a crystal. The properties of the vibrational modes of molecules or atoms in a lattice are largely determined by the mass of the atoms and their bond type. The appearance of any physical factor affecting the short-range order results in modifications in the vibrational characteristics of the atoms and are readily noticeable in the Raman spectrum. This reflects, for instance, in striking differences between the spectrum of an amorphous material and of a crystalline sample of the same kind.

In the setup we used, a laser of 532 nm wavelength was focused to a spot of a few µm. The chosen wavelength probes a thickness of ~1.3 µm [110] under the crystal surfaces. The laser light interacts with atomic vibrations, phonons or other excitations in the system, resulting in the energy of the laser photons being shifted. The shift in energy gives information about the vibrational modes in the system. Results of the characterizations, reported in Figure 7(b) highlight the presence of only crystalline phase of silicon, manifesting with peaks at the ~303 and ~520 cm$^{-1}$ and the absence of other crystalline phases or strained structures, which would manifest as additional peaks in the recorded spectrum [108, 109].

To further investigate the crystalline quality, we also used Rutherford Back-Scattering in channeling condition (c-RBS) to get information related to the atomic ordering of the first atomic layers of both the faces of the crystal which are parallel to the beam and to the beam entry face. c-RBS was carried out using 2.0 MeV $^4$He$^+$ at scattering angle of 160° in IBM geometry at INFN-Legnaro laboratories. Surface $\chi_{min}$, defined as the ratio of the RBS yield under channeling alignment and the yield in random orientation extrapolated at the surface channel, was chosen as quantitative parameter. The higher the degree of crystalline order in the lattice, the lower the value of $\chi_{min}$ on the surface due to the reduction of dechanneling from the defects in the crystal. c-RBS provides information on the presence of crystalline defects up to a depth of ~2 µm with a few nanometer resolution in the crystal depth. Both the surfaces recorded a value for $\chi_{min}$ that is compatible with the value recorded on a crystal with a surface free from crystallographic defects. Moreover, simulations allowed to quantify the surface peak areal density of Si atoms that do not contribute to channeling. It is worth to notice that even for perfect crystals the Si atomic areal density corresponding to this peak is $1*10^{16}$ atoms/cm$^2$ [111, 112]. This is because at this energy the beam focusing due to channeling needs to cross a certain crystal thickness before starting to decrease the backscattering yield. Mapping the crystal surface on both beam entrance and lateral faces with $^4$He$^+$ spots 1x1 mm$^2$ gave surface peak values from $1*10^{16}$ to $1.4*10^{16}$ atoms/cm$^2$. This means that a non-crystalline surface Si fraction can be present as low as from 0 to 0.8 nm over the sample surfaces, reasonably in the form of a SiO$_2$ nano-layer that naturally forms after air exposure [111, 112]

The high order of crystalline perfection was confirmed also by High Resolution Transmission Electron Microscopy (HRTEM). That is a technique where a beam of electrons is transmitted through a specimen to form an image of the crystallographic structure of a sample at an atomic level. Also in this case, characterizations highlight an ordered arrangement of atomic columns preserved up to the crystal surfaces (see figure 8c-d).

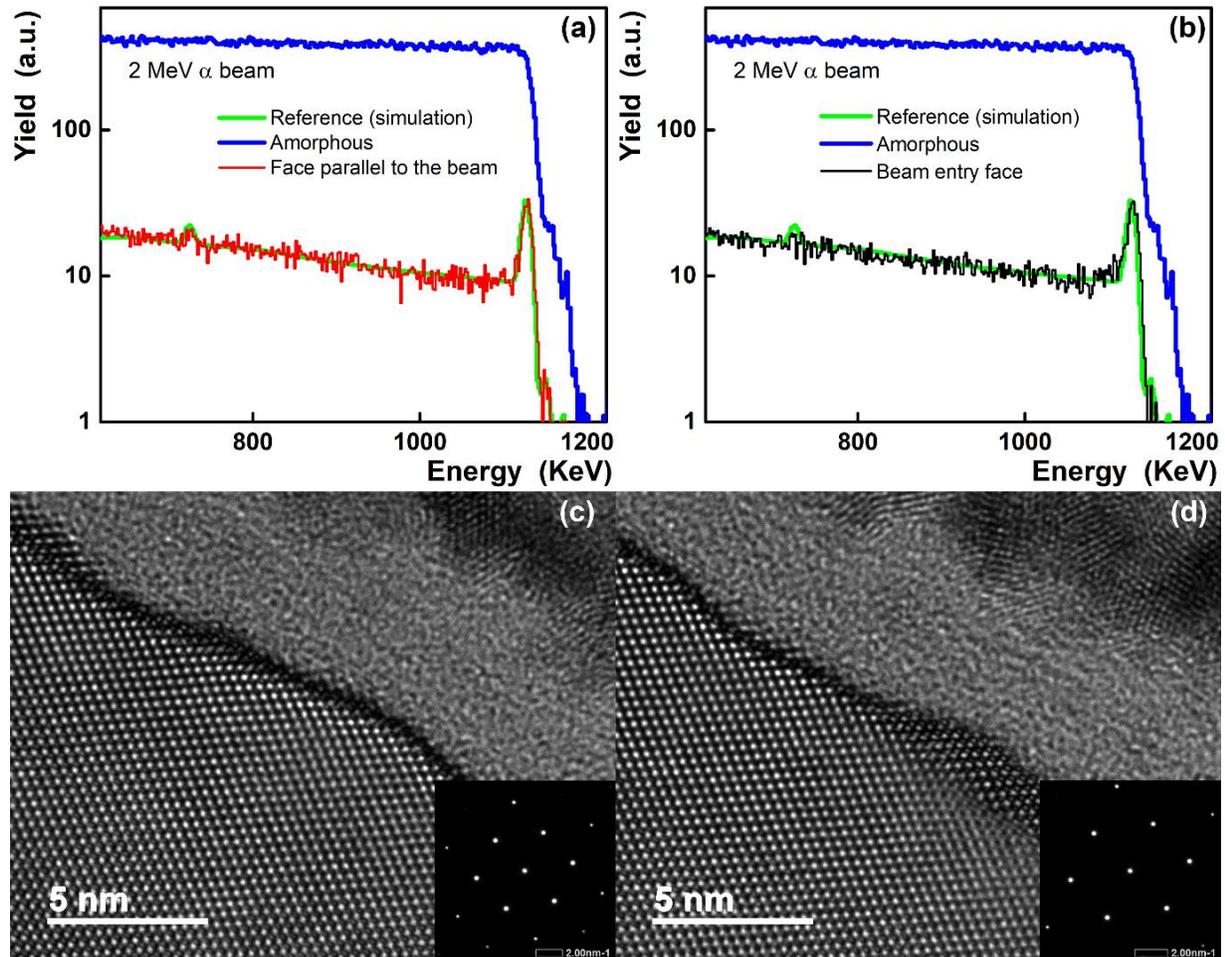

Figure 8. (a) and (b): c-RBS spectra of 2 MeV $^4$He$^+$ recorded respectively on the crystal face parallel and perpendicular to the direction of propagation of a beam in a circular accelerator. (c) and (d): TEM characterizations for the same surfaces. The insets show the electron diffraction pattern of the sample. Both surfaces were found to be free from sub-surface lattice damage induced by MRF and chemical-mechanical polishing operations, respectively.

## 4. Crystal deformation and characterization.

4.1. Control of crystal elastic deformation

The largest scientific field employing bent crystals is X-ray diffraction. The state of the art in controlled deformation of silicon crystals is nowadays dictated by technologies employed worldwide at synchrotrons, where controlled deformation of crystals, typically Si or Ge crystals, is routinely accomplished through mechanical benders actuated by motors. This choice allows compensating mechanical imperfections which arise in mechanical manufacturing of the benders or of the crystals and delivers crystals with deformational field typically controlled within a few nm at the region of interaction between the crystal and the X-ray beam. On the other hand, for the case of the most modern particle accelerators, the usage of motor-actuated benders might represent a relevant risk. Failure of a motor might indeed compromise a key-component of a setup, and its replacement would be an operation far from trivial, as the crystal and its bender would operate in an area not easily accessible, given the high radioactive environment. The material of the bender must be compatible for operations in a ultra-high vacuum environment and should not lead to electron cloud activity in the accelerator [113, 114]. Moreover, the assembly must be as light as possible: at ultra-high energies, such as for the case of the LHC or future accelerators, the critical angle for channeling scales down to values as low as a few µrad or less and as a consequence the crystal must be aligned to the beam with sub-µrad accuracy. The most reliable technology capable of delivering a so extreme accuracy is based on the use of piezo-actuated goniometers [115], whose performances (as for any piezo-actuated motor) are maximized as they operates under the lowest possible weight. This is an additional reason to avoid the use of a motor-actuated bender, as it would likely result in a bulky device. Adoption of a static bender capable of imparting the correct deformational state to the crystal is often the most promising approach, even if mechanical tolerances are often extremely challenging.

A material satisfying the briefly listed requirements is titanium "grade 5", an alloy composed of 90% titanium, 6% aluminum, 4% vanadium. This material, thanks to its low density and extremely high strength, typically finds applications for aerospace, naval and biomechanical applications, engine components, sport equipment, but is very rarely used in ultra-high precision mechanics due to its poor machinability [116].

For the case of crystals with thickness of a few cm and radius of a few meters, the possibility to bend them through clamping of the crystal between properly shaped surfaces of a jaw (see figure 9) is under study. Crystals must operate at an energy range in the TeV region and this requires an extremely high uniformity of the bending radius along the crystal: control of deformational state of the crystal relies on perfection of machining of the surfaces of the bender in contact with the crystal and the surfaces of the crystal itself. That is an approach already developed for extraction of the beam circulating in the SPS [117], which needs to be refined to improve the uniformity of crystal deformations.

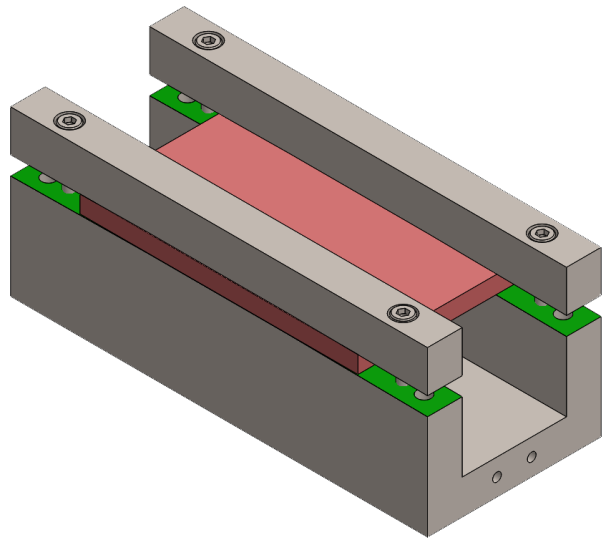

Figure 9: representation of a bent crystal with a geometry suitable for fixed-target experiments in the LHC. The crystal (red part) is clamped between surfaces of a bender with surfaces (green) machined to a cylindrical shape.

For the case of crystals under preparation for studies of the collimation of the LHC circulating beam, we see crystals exploiting anticlastic deformation as the most promising. This approach were initially developed at the Institute for High Energy Physics (Protvino, Russia) [26], and further refined at the INFN-Ferrara (Ferrara, Italy).

Benders are based on revisiting technologies employed for the preparation of bent crystals for the steering of high-energy protons or ions circulating in the SPS [29-31, 83, 118], its extracted lines [119-121] and Tevatron [3], and initially developed at the Institute for High Energy Physics (Protvino, Russia) [26]. Differently from the benders of previous generations manufactured at INFN-Ferrara, which were made of aluminum alloys [122] for an easy machinability, benders are now manufactured in titanium alloys and mechanisms allowing for adjustment of the crystal deformational state in terms of torsion [123] and bending radius are removed, resulting in a geometrically simpler and a lighter support.

Reaching the correct deformational state of the crystal purely relies on ultra-high precision machining processes of the surfaces in contact with the crystal: with the aim to exploit "anticlastic deformation", a "c-shaped" bender was manufactured (see figure 10), where surfaces in contact with the crystal must have a very well controlled inclination. The requirement on bending angle of 50±2.5 µrad translates in a tolerance for the inclination around the *y*-axis of only 0.0022°.

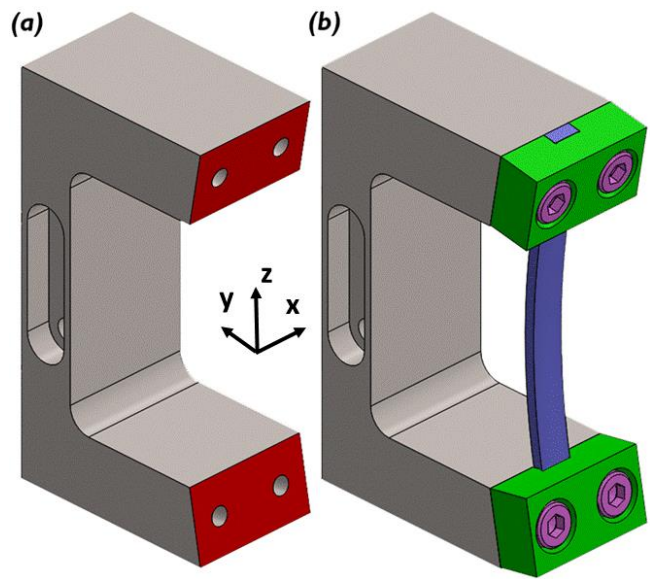

Figure 10 (a) Base structure of a static bender for strip crystals. Red colored surfaces are tilted by 89.9529±0.0022° around the *y*-axis toward the inner side of the bender to impart a "primary bending". The same surfaces are tilted of less than 0.0020° around the *z*-axis to avoid torsion (in this picture the tilting around the *y*-axis is largely increased). (b) Crystal (blue color) assembled on the bender device (deformation of the crystal is exaggerated on purpose in the picture). A couple of clips (green color) clamps the crystal on the holder. Screws (violet color) secures the clamps to the bender.

Aside from imparting a proper bending to the crystal, the bender must avoid generation of unwanted deformations, such as torsional effects. To illustrate the role of torsion for simplicity we refer to the scheme, already described, exploiting anticlastic deformation (see figure 11); similar considerations applies to other possible geometries. The presence of crystal torsion introduces a rate of change of the optimal angle of crystal-beam alignment for unit displacement along the vertical direction of the crystal (see figure11). This reduces the geometrical acceptance of the crystal and its steering efficiency, resulting in a degradation of its performances as it operates in a circular accelerator [30, 31].

For the case of crystal/benders assemblies for collimation of the LHC, crystal torsion might arise as consequence of mechanical imperfections of the holder or inaccuracies in the procedure of assembly of the crystal on the holder. Referring to figure 10, the bender must be machined with approaches delivering small mutual rotations around the *z*-axis of the surfaces in contact with the crystal: thanks to recent developments, a tolerance of 0.0020° (leading to a torsion of 1 µrad/mm) is nowadays reachable and if needed, lower values for torsion can be reached through further developments in holder machining or using benders exploiting hinged flexure mechanisms specifically designed to reduce torsion [26, 123]

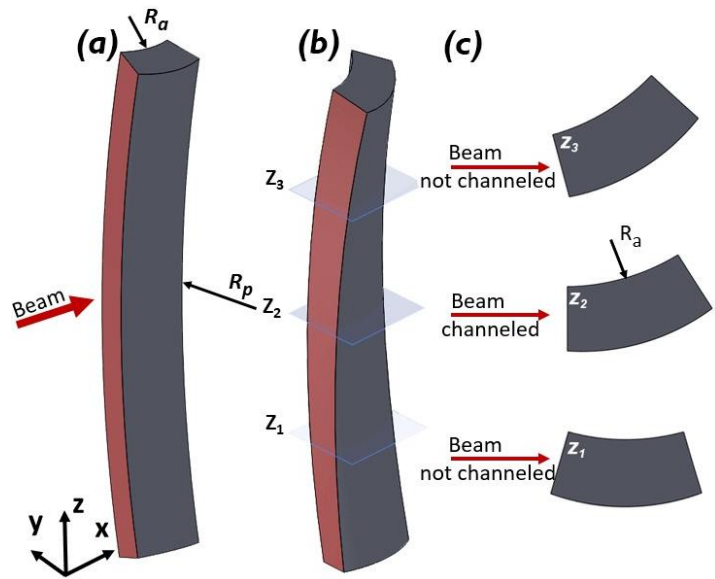

Figure 11 (a) A bent silicon strip crystal. A couple of moment are applied at end of the crystal, resulting in a "primary" bending of the crystal, with radius $R_p$, and a "anticlastic deformation" of crystal cross section, manifesting with a curvature of radius $R_a$. (b) As consequence of mechanical imperfections of the holder or of the mounting procedures, the bent crystal might be subject to a torsion. (c) Cross sections of a bent strip subjected to torsion, taken at three different *z* positions. As a result of the torsion, alignment between crystal cross section and the beam linearly changes along the vertical direction of the crystal, reducing its geometrical acceptance.

4.2. Characterization of crystal deformational state

Prior to installation in an accelerator, the crystal must be properly characterized in terms of its deformational state: tolerances in terms of bending and torsion are indeed challenging to achieve, and inaccuracies in the machining of benders or in the mounting procedure may result in an unwanted deformational state of the crystal.

A preliminary characterization of the deformational state of the crystal was performed using a white-light interferometer (Veeco NT1100). This characterization delivered precise information related to the surface of the crystal. In principle, such characterization could be related to the deformational state of the atomic planes, nevertheless we prefer to perform directed characterizations of the deformational state of atomic planes using a high-resolution X-ray diffractometer. The instrument was interfaced to a custom-made autocollimator to increase its accuracy and precision in a wide class of measurements, among which the characterization of the crystal deformational state. Measurements of bending angle were achieved by mounting the crystal on the Eulerian cradle of the diffractometer: six degrees of freedom allow the alignment of the crystal to the X-ray beam. Figure 12 sketches the method for the measurement of crystal bending angle. The crystal is oriented to the beam in order to obey the Bragg reflection condition from the bent lattice planes at its central region ($X_2$). Deformations of the crystalline planes break the translational symmetry of the system along the $X$ axis, establishing a linear relationship between the position along the $X$ axis and the angle at which the crystal must be oriented to the beam in order to match the Bragg reflection.

Characterization of deformational state of the crystal is accomplished translating the crystal with respect to the beam along the $X$ axis and recording an X-ray rocking curve at each position. The center of mass of each rocking curve is calculated and a map relating centroid of each rocking curve to the $X$ position is generated (inset of figure 12 (b)), delivering exact information on the deformational state of the crystal.

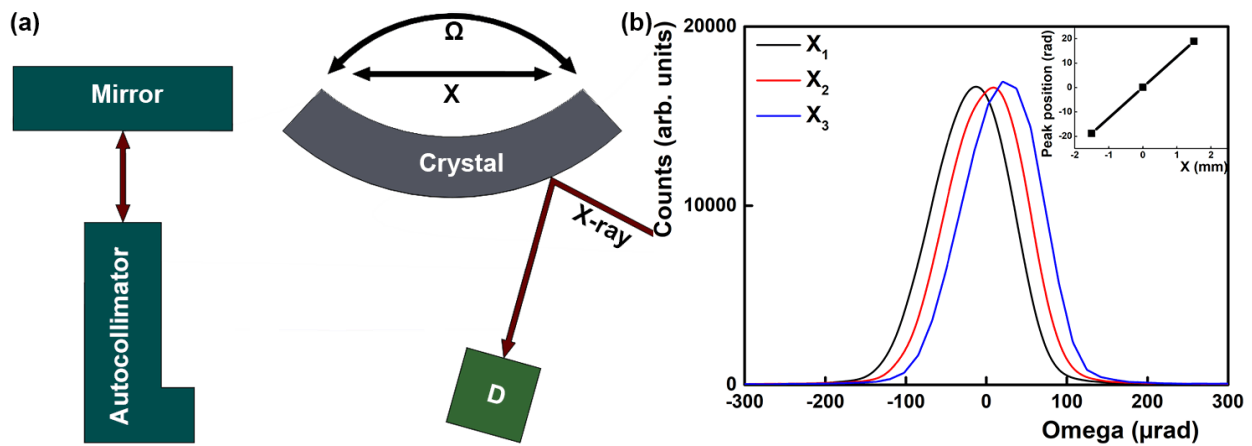

Figure 12 (a) Sketch of the setup employed to characterize the deformational state of the crystal in terms of the bending of the lattice planes. A small portion of the crystal is illuminated with a collimated and monochromatic X-ray beam. The crystal is oriented with respect to the beam in order to match the Bragg reflection condition from its bent planes. The high order reflection (440) is chosen to minimize footprint of the X-ray beam on the crystal: Bragg reflection occurs as the angle between lattice planes and X-ray beam is 53.3508°. The rotation of the crystal around the $\Omega$ axis produces a "rocking curve" (see profiles in (b)). The bending of the crystal breaks the translational symmetry of the system along the X-axis: if the crystal is aligned to the beam in order to match Bragg reflection, alignment is lost after translation of the crystal. Angular motions occurring while rotating the crystal or as result of mechanical tolerance occurring during the translation along the *X* -axis are characterized by means of an autocollimator tracking a mirror integrated to the crystal. (b) Recording a set of rocking curves at various positions while translating the crystal along the *X* axis delivers a map of the crystal deformational state. The center of mass of each rocking curve is calculated and plotted against the corresponding position along the *X* axis: the slope of the linear relationship between the *X* position and the center of mass of the corresponding rocking curves gives a measurement of the bending angle of the crystal.

The measuring setup is strongly influenced by precision and accuracy of the angular motions. With the aim to precisely characterize angular motion of the crystal during rotations around the $\Omega$ axis and to characterize parasitic rotations around the same axis occurring due to mechanical tolerance when the crystal is translated, a laser autocollimator tracks angular motion of a mirror integrated with the crystal. Ultimately, accuracy and precision reached by the autocollimator determines angular accuracy and precision in the reconstruction of the deformational state of the crystal. A first-generation setup employed an autocollimator capable of 1 µrad precision and accuracy; more recently this tool has been upgraded to be capable of a 0.5 µrad accuracy and precision.

The measurement setup has been benchmarked in terms of crystal bending angle against a wide series of experiments involving the use of bent crystals as elements in the steering of high energy particle beams. Those experiments were performed at H8 and H4 external line of the SPS of CERN. Table 1 reports a comparison of the values of the bending angles determined at our laboratory with those determined at the experimental setups exploiting particle channeling. The comparison highlights the reliability of the characterizations conducted with a laboratory equipment.

| Crystal code | Bending angle X-ray (µrad) | Bending angle particle channeling (µrad) |
|---|---|---|
| STF47 | 33±2 | 34 [124] |
| STF48 | 144±2 | 144 [124] |
| STF49 | 247±3 | 247 [124] |
| STF50 | 142±5 | 139 [124] |
| STF51 | 33±2 | 33 [124] |
| STF70 | 56±2 | 54 [125] |
| STF71 | 60±5 | 61 [125] |
| STF99 | 119±3 | 120±2 [75] |
| STF100 | 67±6 | 63±2 [75] |
| STF101 | 170±6 | 165±2 [75] |
| STF102 | 45±3 | 42±2 [126] |
| STF103 | 52±5 | 55±2 [75] |
| STF105 | 49±3 | 50±2 [75] |
| STF106 | 42±2 | 40±2 [127] |
| STF107 | 55±2 | 55±2 [127] |
| STF114 | 52±3 | 52±2 [128] |

Table I. Comparison between bending angles measured through high-resolution X-ray diffraction and those measured in channeling experiments for a large set of crystals. In all the cases the bending angle measurements with the cited techniques are in a good agreement.

## 5. Crystals thermal stability

For the cases where the crystal must operate under ultra-high vacuum, a key parameter that the bender-crystal assembly must satisfy is the stability of its deformational state with respect to bake-out cycles which are needed to reach the working condition. For the case of LHC, the achievement of the operational conditions was accomplished through a series of bake-out cycles which brings the crystal and the bender at 250 °C for 48h for each cycle [129]

In order to assure thermal stability of the bender material, a series of thermal annealing were performed on the titanium material prior and after its machining process. We assume thermal stability of the assemblies if the crystal does not change its deformational state after at least 10 vacuum bake-out cycles: the crystal deformational state was characterized before and after each thermal cycle.

Figure 13 reports the measured bending angle and torsional values for a crystal subject to a total of 50 thermal cycles, highlighting the robustness of the assembly against these thermal cycles.

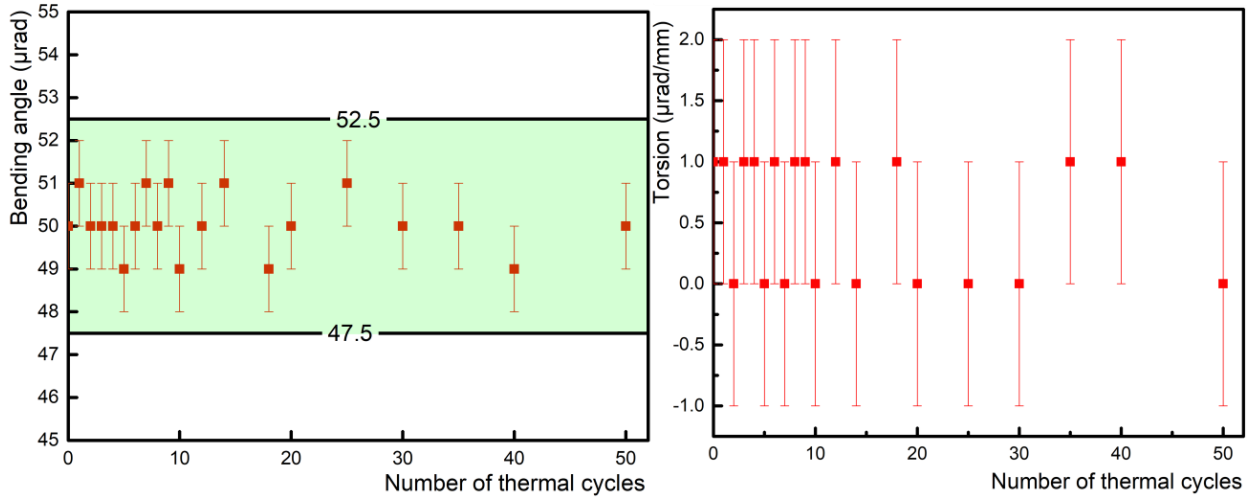

Figure 13. Study of the thermal stability of the crystal against bake-out thermal cycles. (a) Value of bending angle vs the number of thermal cycles. Green area highlight the region of acceptable bending angle of the crystal following [63]. (b) Measured torsion as a function of the number of thermal cycles. The stability of the bending angle and torsion values demonstrates the robustness of the crystal deformation state and of the bender.

## 6. Conclusions

Thanks to their well-ordered atomic structure, crystalline materials have been studied since few decades as elements to steer particle beams at particle accelerators. As a result of considerable technological advances, their applications at ultra-high energy and intensity accelerators is now possible. Such a result is achieved thanks to a highly interdisciplinary approach developed at INFN-Ferrara merging the most advanced techniques typically used in the fields of silicon micromachining, ultra-precise optics manufacturing, X-ray topography and ultra-precise metrology.

The results of laboratory-based investigations highlight the crystallographic perfection of the crystals starting from their first few atomic layers, an ultra-low miscut angle of just few µrad and a reliable correlation between the deformational state of the crystal characterized by means of high-resolution X-ray diffraction and the channeling of high energy particle beams.

After a long lasting experimental investigation carried at CERN mainly in the frame of the UA9 collaboration, this technology has been integrated into the HL-LHC baseline program for the realization of a crystals-based setup for collimation for heavy-ion beams. At the same time, availability of this technology opens new possibilities also for beam extraction and to perform innovative fixed-target experiments at the most advanced particle accelerators.

Besides that, crystals might play a key-role also to achieve slow extraction of particle beams circulating in lower energy accelerators such as the SPS at CERN or at synchrotrons facilities, as innovative sources of gamma radiation, and as elements useful for the production of positron or beams to be used in a wide class of physics investigations.

**Acknowledgments**

We acknowledge Professor Lucio Rossi and Dr. Walter Scandale for their long-lasting support to this activity in the frame of the UA9 collaboration. A. Mazzolari acknowledge Professor Lucio Rossi, Dr. Stefano Redaelli, Dr. Simone Gilardoni, Dr. Alessandro Masi, Dr. Alexander Taratin, Professor Victor Tikhomirov, Dr. Valery Biryukov and Professor Geoffrey Hall for many stimulating conversations on the subjects treated in the manuscript and for critically reading it. A. Mazzolari acknowledge Andrea Persiani and Claudio Manfredi (PERMAN), Dr. Claudio Sgarbanti (Biomeccanica) and Dr. Riccardo Signorato (CINEL) for their friendly support with manufacturing of crystal benders. This work has been supported by the ERC Consolidator Grant CRYSBEAM GA, the ERC Consolidator Grant SELDOM GA 771642 and by the INFN CSN1. A. Mazzolari and V. Guidi acknowledge founding from PRIN 2015LYYXA8 "Multi-scale mechanical models for the design and optimization of micro-structured smart materials and metamaterials".